\documentclass[aps,apl,showpacs,twocolumn,superscriptaddress,longbibliography]{revtex4-1}
\usepackage[pdftex]{graphicx}
\usepackage[pdftex,bookmarks,colorlinks,breaklinks]{hyperref} 
\usepackage{amsmath,amssymb}
\hypersetup{linkcolor=red,citecolor=blue,filecolor=dullmagenta,urlcolor=blue} 
\hypersetup{linkcolor=red,citecolor=blue,filecolor=blue,urlcolor=blue} 
\hypersetup{pdftoolbar=true,pdfpagemode=UseNone,pdfstartview=FitH,colorlinks=true,extension=}

\renewcommand{\eqref}[1]{(\ref{#1})}
\newcommand{\Ind}{\ensuremath{I_{n_D}}}
\newcommand{\OInd}{\ensuremath{\Omega(\Ind)}}

\newcommand{\C}{\ensuremath{{\cal{C}}}}

\renewcommand{\Re}{\ensuremath{{\textrm{Re}}}}
\renewcommand{\Im}{\ensuremath{{\textrm{Im}}}}

\begin{document}
\title{Self-Shielded Topological Receiver Protectors}
\author{Mattis Reisner}
\thanks{These two authors contributed equally}
\affiliation{Universit\'{e} C\^{o}te d'Azur, CNRS, Institut de Physique de Nice (INPHYNI), 06108 Nice, France, EU}
\author{Do Hyeok Jeon}
\thanks{These two authors contributed equally}
\affiliation{Wave Transport in Complex Systems Lab, Department of Physics, Wesleyan University, Middletown CT-06459, USA}
\author{Carsten Schindler}
\affiliation{Fachbereich Physik, Philipps Universit\"{a}t Marburg, 35032 Marburg, Germany}
\author{Henning Schomerus}
\affiliation{Department of Physics, Lancaster University, Lancaster, LA1 4YB, United Kingdom}
\author{Fabrice Mortessagne}
\affiliation{Universit\'{e} C\^{o}te d'Azur, CNRS, Institut de Physique de Nice (INPHYNI), 06108 Nice, France, EU}
\author{Ulrich Kuhl}\email{ulrich.kuhl@univ-cotedazur.fr}
\affiliation{Universit\'{e} C\^{o}te d'Azur, CNRS, Institut de Physique de Nice (INPHYNI), 06108 Nice, France, EU}
\affiliation{Fachbereich Physik, Philipps Universit\"{a}t Marburg, 35032 Marburg, Germany}
\author{Tsampikos Kottos}\email{tkottos@wesleyan.edu}
\affiliation{Wave Transport in Complex Systems Lab, Department of Physics, Wesleyan University, Middletown CT-06459, USA}

\begin{abstract}
Receiver protectors (RPs) shield sensitive electronics from high-power incoming signals that might damage them. Typical RP schemes range from simple fusing and PIN diodes, to superconducting circuits and plasma cells 
-- each having a variety of drawbacks ranging from unacceptable system downtime and self-destruction to significant insertion losses and power consumption. 
Here, we theoretically propose and experimentally demonstrate a unique self-shielding RP based on a coupled-resonator-microwave-waveguide (CRMW) with a topological defect being inductively coupled to a diode. 
This RP utilizes a charge-conjugation (\C) symmetric resonant defect mode that is robust against disorder and demonstrates high transmittance at low incident powers. 
When incident power exceeds a critical value, a self-induced resonant trapping effect occurs leading to a dramatic suppression of transmittance and a simultaneous increase of the reflectance close to unity. 
The proposed RP device is self-protected from overheating and electrical breakdown and can be utilized in radars, reflection altimeters, and a broad range of communication systems.
\end{abstract}


\maketitle

\section{INTRODUCTION}

\begin{figure}
	\centering
	\includegraphics[width=.99\columnwidth]{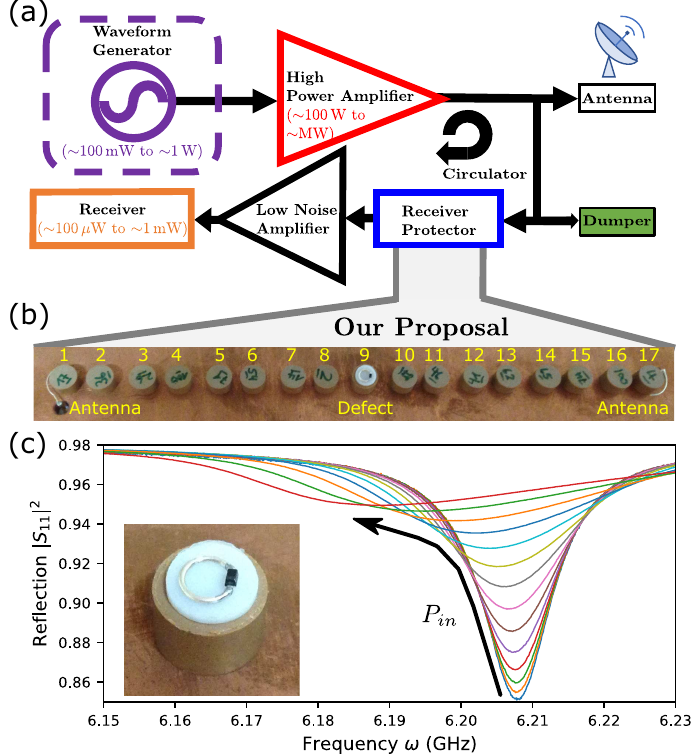}
	\caption{\label{fig:fig1}
		The proposed implementation of a RP based on self-induced spontaneous \C-symmetry violation of a topological defect state.
		(a) An application in a radar setup.
		(b) The RP, consisting of 17 dielectric resonators with complex eigenfrequencies $\nu_n=6.885$\,GHz$+i 1.7$\,MHz structured by four dimers to the left, the central defect, and four dimers to the right.
		The incoming signal is injected via the left antenna with a power $P_{in}$ and transmitted via the right antenna.
		The central resonator is weakly coupled to both sides and has a diode positioned on top of it using a Teflon spacer of 1-mm height [see photograph in (c)].
		(c) The reflection spectra $R=|S_{11}|^2$ for an isolated resonator inductively coupled to a diode for increasing input power $P_{in}$ ranging from -30 (blue) to 4.5\,dBm (red) in steps of 2.5\,dBm. 
		The black arrow indicates the direction of increasing power.
		For small powers a decrease of the resonance height due to an increase of the resonance width is observed, whereas for higher powers an additional frequency shift is seen as well.
		To obtain these spectra a single antenna is weakly coupled to the system ($t_{L}=16.5$\,MHz).
	}
\end{figure}

The Internet of Things, where everyday devices and objects communicate with each other via electromagnetic waves, is an incipient revolution that will dramatically affect future generations. 
The envisaged intimate exchange of information between autonomous vehicles, drones, remotely controlled robots, and "smart homes" builds on stunning technological developments in controlling electromagnetic signals with unprecedented precision. 
At the heart of the present endeavors are 5G technologies targeting digital cellular networks \cite{transceiver}.
The communication within such a network can be either direct or based on a server. 
In both cases each device needs to have a communication channel, i.e., an antenna that acts as a transceiver \cite{transceiver} (an automated transmitter and receiver) for electromagnetic waves. 
These combined schemes require not only an efficient signal reception, but must also be equipped with a circuit component that will protect their sensitive electronics from undesirable (deliberate or accidental) high-power incident radiation that is threatening to damage them. 
These {\it receiver protector} (RP) schemes should be efficient, cheap, and once the threat from the damaging radiation is surpassed they must restore the transceiver 
to full functionality as fast as possible without any external interaction.

Currently, RPs are already indispensable elements of various microwave systems, such as radars, reflection altimeters, and communication systems, the sensitive electronic components which are susceptible to damage from high-power microwave energy \cite{rob91,RP1,RP2}.
The latter can couple into these systems through intended signal paths (such as antennas or sensors) or unintended entry points (such as enclosure slots and wire harnesses).
In other occasions, such as in simultaneous-transmit-and-receive (STAR) radar systems \cite{transceiver}, the threatening high-power microwave signals may be the reflected remnants of their own radar transmitter output.
These signals are usually on the order of kilowatts to megawatts of peak power -- far too much to survive for any electronic system at the receiver end [see Fig.~\ref{fig:fig1}(a)].
Typical RPs are located between the antenna and the sensitive rf components and are often composed of multiple stages to meet protection specifications \cite{rob91,cho16,RP2}.
In these configurations, a diode limiter \cite{RP2} is used at the final stage of protection.
A serious drawback of this scheme is the significant insertion loss; specifically, if multiple diode stages are utilized.
Another commonly used RP scheme involves gas tubes contained in a piece of the transmission waveguide \cite{cro13,tube}.
The limiting action occurs due to the breakdown of the gas at high powers into a low-impedance plasma.
Such RPs have limited operational lifetime due to direct contact with the plasma, which erodes the electrodes and contaminates or entraps the fill gas \cite{cro13}.
For completeness we also mention superconducting-based RP technologies \cite{boo03,super1,super2,super3}, which, however, require a high power consumption in order to keep the superconducting circuit at low temperatures.
To summarize, the new generation of RPs should:
(a) maintain low insertion losses for low incident powers,
(b) develop abrupt limiting action at the required limiting-threshold (LT) powers and, crucially,
(c) also protect the RP itself from self-destruction due to overheating or electrical breakdown.
Ideally, one also wants to control the LT externally.
The input power (or fluence threshold) above which the RP is destroyed is known as the damage threshold (DT).
The ratio of the DT to LT defines the dynamic range of the limiter and constitutes a figure of merit of the RP performance.
This reconfigurability is extremely useful in the case of multifunction rf system applications that allow many users to share the same antenna.

In this paper we utilize concepts from topological photonics in order to demonstrate a self-shielding microwave RP consisting of a one-dimensional array of resonators coupled to a diode.
This structure simultaneously satisfies many of the above constraints via a unique protection mechanism.
Topological photonics \cite{tom19} aims to implement ideas coming from the concept of a topological insulator \cite{has10,kan11}, originating in the field of electronic transport in condensed matter.
This approach allows us to target devices with special functionalities via specific designs of photonic circuits.
In this endeavor, the implementation of symmetries in the circuits has proven extremely useful.
One particular achievement in these efforts is the realization of symmetry-protected topological defect modes that are resilient to fabrication or environment-induced imperfections \cite{sch13,pol15,mak14,mak15,kuh17,rei19} and their implementation in applications such as topological defect-mode lasers \cite{stj17,zha18,par18}.
In this paper, we utilize a nonlinearly functionalized topological mode as a RP communication gateway that efficiently decouples at the LT, thereby reflecting harmful signals while safeguarding both the downstream components and the RP elements themselves.

\section{PRINCIPLE OF OPERATION OF THE RECEIVER PROTECTORS}
The basic idea of the proposed RP is as follows.
For low incident powers, the RP shows a high peak resonant transmission associated with a topologically induced resonant-defect mode, based on a charge-conjugation (\C) symmetry as detailed further below.
When the power of the incident signal exceeds a critical value, the nonlinear losses of the diode are activated.
Then, the resonant mode experiences a self-induced transition from an underdamping to an overdamping regime, leading to a destruction of the resonant-defect mode and a subsequent abrupt suppression of the transmittance.
At the same time, the reflectance acquires high (near-unity) values.
Consequently, the sensitive diode is protected from self-destruction due to heating via the absorption of the incident energy, which is now reflected back to a dumper channel.
At the same time, the destruction of the resonant-defect mode at high incident powers also ensures the protection of the diode from electrical breakdown.
Note that the diode by itself can be used as a RP, as in conventional setups.
Its implementation, however, as a component of the \C-symmetric photonic circuit \cite{sch13,pol15} leads to a prominently enhanced dynamic range.
We demonstrate this point by comparing the performance of the stand-alone diode with that demonstrated by our photonic circuit.
Further insight into the RP is gained by a theoretical modeling of its transport characteristics.
The analysis also allows us to identify the upper limits of the RP performance, which are associated with the bulk absorption and the couplings between the photonic circuit and the antennas.

\section{THE RECEIVER PROTECTOR AND ITS COMPONENTS}
An immediate application of our design is associated with radar STAR systems [see Fig.~\ref{fig:fig1}(a)].
The RP consists of a bipartite coupled resonator microwave waveguide (CRMW) array with a nonlinear lossy defect [see Fig.~\ref{fig:fig1}(b)].
The lattice is implemented using an array of $N=17$ high-index cylindrical resonators [radius $r=4$\,mm, height $h=5$\,mm, made of ceramics (ZrSnTiO) with an index of refraction $n_r\approx 6$] with a resonance frequency around $\nu_0\approx 6.885$\,GHz and a line width $\gamma\approx 1.7$\,MHz.
The resonators are placed at alternating distances $d_1=12$\,mm and $d_2=14$\,mm corresponding to strong ($t_1$) and weak ($t_2$) evanescent couplings, respectively.
The leftmost and rightmost resonators are connected to kink antennas [see Fig.~\ref{fig:fig1}(b)], which are curled around the resonators to guarantee strong coupling to the electric field in the $x-y$ plane.
The structure is shielded from above by a metallic plate (not shown).
A topological defect is introduced by repeating the spacing $d_2$ (weak coupling) around the ninth resonator ($n_D=9$).

Static linear losses in a RP resonant mode can lead to a dramatic suppression of the transmittance and a simultaneous increase of the reflectance and drop of the total absorbance of the structure \cite{kuh17}.
However, such a static realization would miss the crucial ingredient necessary for its implementation as a RP, i.e., a self-regulated (nonlinear) loss mechanism triggered by the intensity of the incoming field \cite{mak14,mak15}.
Indeed, in the microwave regime, there is a lack of materials with considerable nonlinear losses -- as opposed to the optical and IR, where materials with significant two-photon absorption mechanisms are available \cite{tut93,optics2}.

A "hybrid" CRMW-diode structure can bypass this constraint and can be used for the implementation of nonlinear losses in the microwave domain.
The latter are introduced by placing the diode (detector Schottky Diode SMS\,7630-079LF, from Skyworks) above the defect resonator at $n=n_D=9$ using a Teflon spacer [see photograph in Fig.~\ref{fig:fig1}(c)].
The diode is short circuited and coupled via a metallic ring with a diameter of 3\,mm.
Thus the $z$-directional magnetic field \cite{bel13b} at the defect resonator of the transmitted signal is inductively coupled to the fast diode.
The strength of the magnetic field dictates the value of the current at the ring and consequently the voltage across the diode.
The latter dictates the state of the diode:
the “on” state is associated with high voltage (high incident power) and leads to high losses at the defect resonator; the “off” state is associated with low voltage (low incident power) and leads to low losses.

\section{MODELING OF THE PHOTONIC CIRCUIT}
The CRMW array can be described by an effective coupled-mode theory Hamiltonian $H$
\begin{eqnarray}\label{eq:CMTeqs}
\omega\psi_n&=&\sum_mH_{nm}\psi_m,
\\\nonumber
H &=& \sum_n \nu_n |n \rangle \langle n| + \sum_n t_{n,n+1} (|n \rangle \langle n+1| + |n+1 \rangle \langle n|),
\end{eqnarray}
where $\psi_n$ is the magnetic field amplitude and $\nu_n=\nu=\nu_0+i\gamma$ is the resonance frequency of the $n$th resonator.
The coupling constants $t_{n-1,n}$ describe the coupling between the $(n-1)$th and $n$th resonator and take only two values, $t_1$ and $t_2$.
Furthermore, we can show that $H\C+\C H\equiv \{H,\C\}=0$ where $\C \psi_n = (-1)^n \psi_n^*$, i.e., the Hamiltonian $H$ anticommutes with an antilinear operator (involving a complex conjugation), which defines a charge-conjugation symmetry and can lead to topological states \cite{zir97}.

In the absence of the defect resonator, one can invoke Bloch's theorem and derive the dispersion relation
$\omega(k) = \nu_0 \pm \sqrt{t_1^2+t_2^2+2t_1t_2\cos(k)}$ with wave number $k$ (in the case in which $\gamma\ne 0$, the spectrum is ${\tilde \omega}(k)=\omega(k)+i\gamma$),
indicating the existence of two minibands at frequency intervals
$\nu_0 - t_1-t_2 < \omega < \nu_0 - |t_1 - t_2|$ and $\nu_0 + |t_1-t_2| < \omega < \nu_0 + t_1 + t_2$
separated by a band gap of width $2|t_1-t_2|$.
By means of transmission measurements, we determine the widths of the minibands and the gap, which allows us to extract the coupling values $t_1=68$\,MHz, $t_2=33$\,MHz.
These are in agreement with the values obtained from the resonance splitting observed for two coupled resonators with the corresponding distances $d_1$ and $d_2$ \cite{bel13b,bar13a}.
The defect created by the two consecutive weak couplings supports a topological defect state with frequency $\nu_0$ in the middle of the gap.
The mode profile has a characteristic staggered shape given by \cite{pol15}
\begin{equation}\label{eq:dmode}
\psi^D_n \propto
\left\{	
\begin{array}{ll}
\frac{(-1)^{\frac{|n-n_D|}{2}}}{\sqrt{\xi}}e^{-\frac{|n-n_D|}{2\xi}}, & \textrm{if } |n-n_D| \textrm{ is even,}\\
0, & \textrm{if } |n-n_D| \textrm{ is odd,}
\end{array}
\right.
\end{equation}
where $\psi^D_n$ is the wave function of the defect state at the $n$th resonator site.
The localization length $\xi$ is given by $\xi = 1/\ln(t_1/t_2)$.
Note that $n$ refers to site indices for each individual resonator.
Given that $n_D$ is odd, the stationary solution is supported only by the odd-$n$ sublattices. This mode is protected by the \C symmetry against positional disorder, even in the presence of inhomogeneous losses, as long as the latter respects the bipartite nature of the lattice \cite{sch13,pol15}.

When the diode is coupled to the defect resonator $n_D=9$, it alters its complex resonant frequency $\nu_{n_D}=\nu_0 + \OInd$, where $\Ind = |\psi_{n_D}|^2$ is the local field intensity.
The functional form of the PIN-diode nonlinearity matches well with a saturable nonlinear absorption
\begin{equation}\label{eq:Omega}
\OInd = z_0 - z_1 \frac{1}{1+\alpha \Ind}\,,
\end{equation}
which is extracted by analyzing, both theoretically and experimentally, the setup consisting only of a single antenna directly coupled to an isolated resonator, which is inductively coupled to the diode [see Fig.~\ref{fig:fig1}(b) and (c)].
Using the coupled-mode theory for this setup, we evaluate analytically the reflectance $R$ and absorbance $A$ and fit these expressions to our measurements for various incident powers $P_{in}$ in order to extract the fitting constants $z_0=(-50+29.6i)$\,MHz, $z_1=(-50+22i)$\,MHz, and $\alpha= (0.2-2i)$\,mW$^{-1}$.
For the two limiting cases of weak and strong field intensities, we obtain $\OInd= z_0 - z_1$ for $(I\ll 1)$ and $\OInd= z_0$ for $( I\gg 1)$, respectively.
For low incident powers, $\OInd= z_0 - z_1 \approx 7i$\,MHz is purely imaginary.
At the other limiting case of high incident powers, the nonlinear term reduces to $\OInd= z_0$, with a real part, that is $\Re(z_0)\ll \nu_0$ (by 2 orders of magnitude) and thus can be safely disregard for our modeling.
In fact, the same argument can be generalized for the case of moderate incident powers, where one has to take into consideration the full expression for $\OInd$ given in Eq.~\eqref{eq:Omega}.
On the other hand, the resonance line width is substantially increased as a function of the incident power.
For example, for high incident powers we have that $\Im(z_0)\gg \Im(\nu)$ by an order of magnitude.
We therefore conclude that the increase of the power of the incident radiation mainly affects the imaginary part of the resonant defect frequency $\nu_{n_D}$.

Finally, in the case of the scattering setup, the left and right antennas are modeled as ideal leads, corresponding to semi-infinite tight binding chains with a constant hopping amplitude $t_0\approx 100$\,MHz and a constant on-site potential $\epsilon_n\approx \nu_0$.
Such leads support propagating waves with the dispersion relation $\omega = \nu_0+2t_0\cos(k)$.
The coupling coefficients $t_L\approx 113.5$\,MHz ($t_R\approx 90.5$\,MHz) between each antenna and the leftmost (rightmost) CRMW resonator are generally different from $t_0$ and do not depend on $P_{in}$.
Their values are determined by a least-square fit of the measured transmission spectrum $T(\omega)$ with the corresponding calculated one using our modeling scheme.

\begin{figure}
	\centering\includegraphics[width=\linewidth]{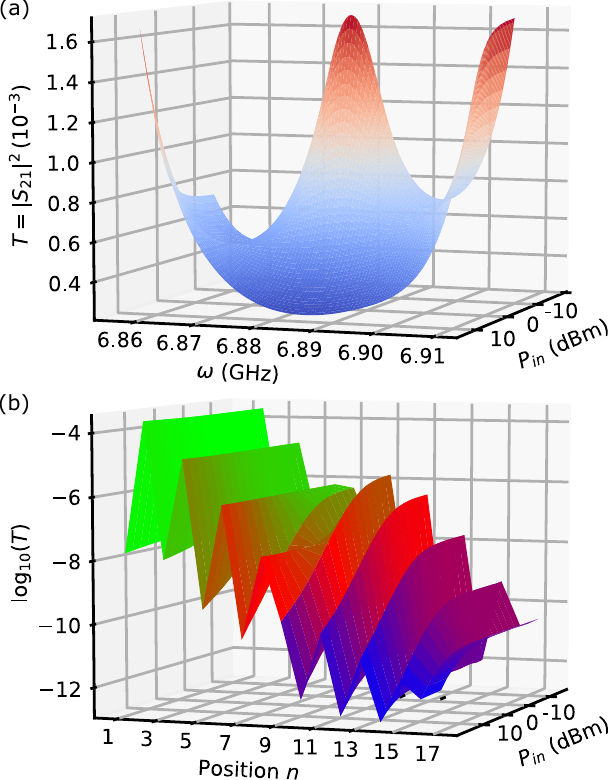}
	\caption{\label{fig:fig2_Exp}
		(a) The measured transmission $T=|S_{12}|^2$ as a function of the frequency $\omega$ and the input power $P_{in}$.
		(b) The logarithm of the transmittance $\log_{10}(T_n)$ at resonator $n$ as a function of the input power $P_{in}$.
	}
\end{figure}

\section{SCATTERING ANALYSIS}
The measured transmittance $T$ close to the resonant-mode frequency $\nu_0$ at the middle of the band gap is reported in Fig.~\ref{fig:fig2_Exp}(a).
The peak observed for small incident powers $P_{in}$ is suppressed with increasing $P_{in}$.
The increase at the edges are due to the bands generated by the bipartite structure.
We further check that the spectral position of the peak is not affected by a positional randomness of the resonators (as long as they preserve the bipartite nature of the structure);
a consequence of the topological protection due to \C symmetry.

In the previous section, we discussed the infinite Su-Schrieffer-Heeger (SSH) model and detail the local nonlinearity placed at the defect site $n_D$.
To analyze the transport properties for the finite CRMW array theoretically, we now turn to the backward-transfer-matrix approach \cite{TH99,MKT19}.
The one-step backward transfer matrix $M_n$ is given by
\begin{equation}\label{eq:Mn}
\left(\begin{array}{c} \psi_{n-1}\\ \psi_{n}\end{array}\right)=
M_n \left(\begin{array}{c} \psi_{n}\\ \psi_{n+1}\end{array}\right),
\qquad
M_n=
\left(\begin{array}{cc}
\frac{\omega-\nu_n}{t_{n-1}} & - \frac{t_{n+1}}{t_{n-1}}\\
1 & 0
\end{array}\right),
\end{equation}
where $\psi_n$ is the wave function of the scattering mode at the $n$th site.
We supplement Eq.~\eqref{eq:Mn} with the appropriate scattering boundary conditions, $\psi_n=te^{i n k}$ for $n\ge N+1$ and
$\psi_n=r_0e^{i n k}+ r e^{-i n k}$ for $n \le 0$ describing a left-incident propagating wave with amplitude $r_0$ and a reflection coefficient $r$.
Using the backward-transfer-map approach, we obtain the transmittance and the scattering profile of the defect mode.

\begin{figure*}
	\centering
	\includegraphics[width=.8\linewidth]{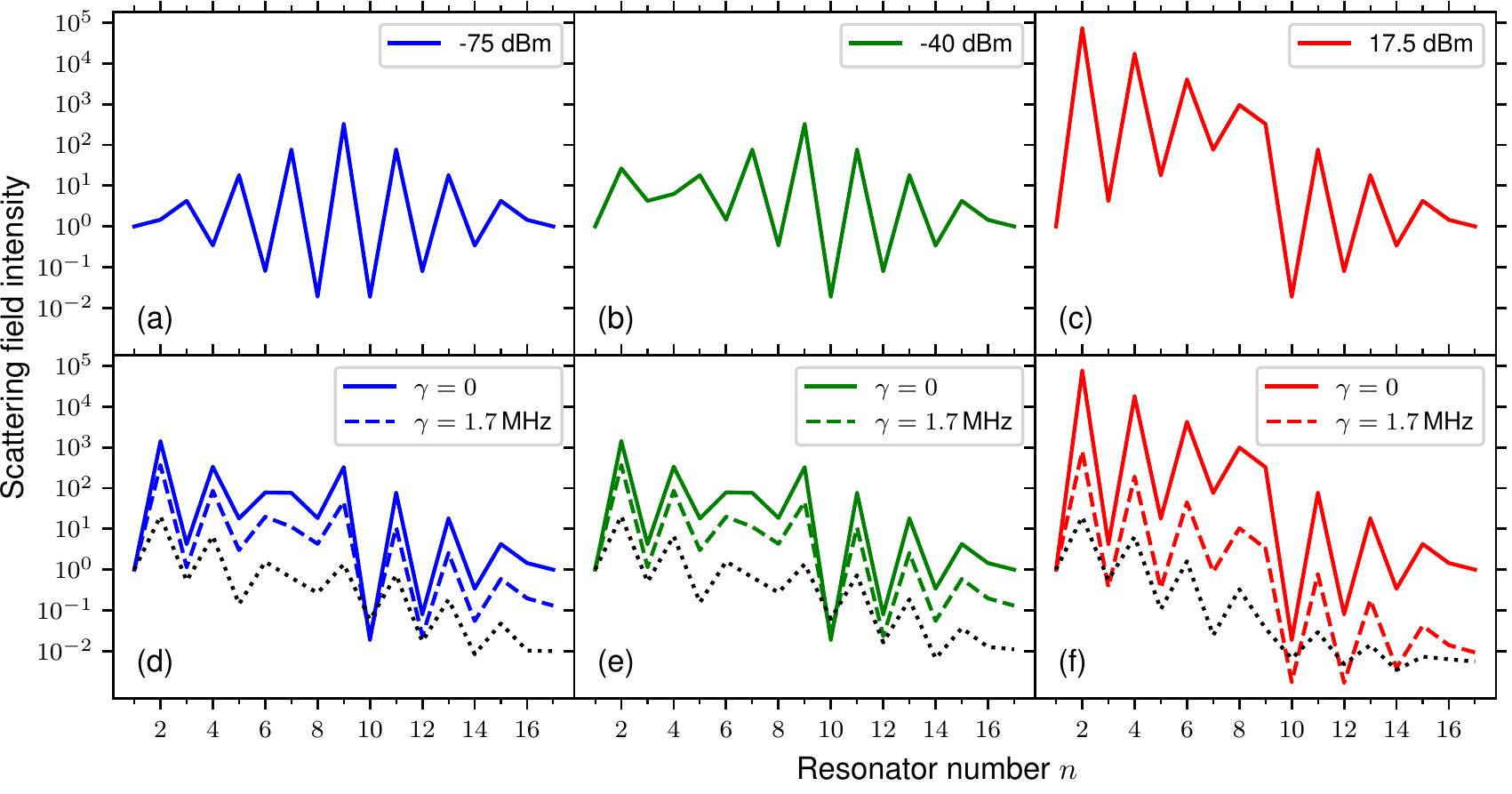}
	\caption{\label{fig:fig3}
		The scattering profile for three representative incident powers:
		(a),(d) weak incident power (-75\,dBm);
		(b),(e) intermediate incident power (-40\,dBm);
		(c),(f) high incident power (+17.5\,dBm).
		In (a)-(c), the bulk losses are set to zero ($\gamma = 0$), an (approximately) symmetric coupling is assumed ($t_L=113.5$\,MHz $\approx t_R=90.5$\,MHz), and the defect resonance is perfectly adjusted ($\Omega(I) = 0$, i.e., $z_0 = z_1$) for low incident powers.
		In (d)-(f) we assume that $z_0 = (-50+29.6i)$\,MHz and $z_1 = (-50+22i)$\,MHz take the values extracted from the best fit of transport measurements of the diode.
		The solid lines correspond to the ideal case in which the bulk losses at each resonator are $\gamma = 0$, the dashed lines to $\gamma = 1.7$\,MHz (the measured experimental value of our resonators), and the dotted line to the experimentally obtained profile. 
		In all cases, the scattering profiles are normalized with respect to the intensity at the first resonator $|\psi_1|^2$.
	}
\end{figure*}

The transport features of our RP can be quantified via a direct numerical evaluation of the transmittance $T = |t/r_0|^2$, reflectance $R = |r/r_0|^2$, and absorbance $A = 1-T-R$ using the
backward map associated with Eq.~\eqref{eq:Mn}.
A further theoretical analysis can be carried out by noting that the transport process is divided into different parts:
propagation before or after the nonlinear defect resonator and propagation through the nonlinear defect.
The first two (essentially identical) processes can be easily carried over using the transfer matrix (\ref{eq:Mn}).
An analytical expression for the total left or right backward transfer matrix $M^{(L/R)}$ is derived using Chebyshev's identity.
Using $M^{(L/R)}$, we calculate the field amplitudes just before and after the nonlinear defect.
Substituting these expressions into Eq.~\eqref{eq:Mn} for $n=n_D=9$, we derive final expressions for the transmittance and reflectance as follows:
\begin{equation}
T=\frac{1}{|\mu_1|^2\left|\left[\left(\frac{\omega-\{\nu+\Omega(\left|t\chi_1\right|^2)\}}{t_2}\right)+\frac{\mu_2}{\mu_1}\right]\chi_1-\chi_2\right|^2}
\end{equation}
and
\begin{equation}
R=T\left|\left(\mu_1^*\frac{\omega-\{\nu+\Omega(\left|t\chi_1\right|^2)\}}{t_2}+\mu_2^*\right)\chi_1-\mu_1^*\chi_2\right|^2\,,
\label{Rexp}
\end{equation}
where $\mu_j=\frac{i}{2}M_{1j}^{(L)} \textrm{csc}\,k + \frac{1}{2}M_{2j}^{(L)} (1-i \cot k)$ and $\chi_j=M_{j1}^{(R)}+M_{j2}^{(R)}e^{ik}$.
In the above expressions, we assume lossless resonators, i.e., $\gamma= 0$.
In the case of $\gamma\ne 0$, the conjugation symbol `$^*$' in Eq.~\eqref{Rexp} has to be understood so that it applies only to the imaginary unit and not to the matrix elements $M_{1j}^{(L)}$ or $M_{2j}^{(L)}$ which, in this case, are also complex numbers.

\section{RESONANT FIELD PROFILE AND PERFORMANCE OF RECEIVER PROTECTORS}
In Fig.~\ref{fig:fig2_Exp}(b), we report the transmittance $T_n$ from the incident antenna to a scanning antenna placed on top of the $n$th resonator.
In the weak coupling limit between the scanning antenna and the resonator, the measured transmittances are directly related to the field intensities at the $n$th resonator. 
The high transmission on the incident side ($n=1$) is defined by the decay of the evanescent coupling to the incident four dimers.
For small incident powers, an increase is seen at the $A$ (odd) sites, leading to higher transmission through the RP. 
Note the switching of the high transmission values from $B$ (even) to $A$ (odd) sites around the defect site ($n_D=9$) for weak incident powers. 
At $P_{in}\approx -5$\,dBm a decrease of $T_n$ is observed at the defect position $n_D$ and the transport for larger $P_{in}$ is then dictated by sites with even index $n$, i.e., the transport is governed by an evanescent wave in the band gap. 
For large incident powers, the support for transmission is mainly from $B$ sites, i.e., due to the evanescent wave of a pure-dimer chain with eight dimers.

In Fig.~\ref{fig:fig3}, we report the numerically evaluated scattering profile for fixed incident frequency $\omega=\nu_0$ and three representative incident field intensities.
We first consider (upper row of Fig.~\ref{fig:fig3}) the simple case in which $z_0 = z_1$ and $\gamma=0$. 
In the limit of weak incident powers $\Omega(\Ind)= 0$ [see Fig.~\ref{fig:fig3}(a)], the mode is exponentially localized around the defect resonator and bears the characteristic signatures of the staggered form associated with the defect mode of the corresponding isolated \C-symmetric Hamiltonian [see Eq.~\eqref{eq:dmode}].
One difference is that the nodes of the defect mode [Eq.~\eqref{eq:dmode}] turn to quasinodal points. We trace this feature back to the existence of a small (but finite) line-width broadening of the resonant mode due to the coupling with the antennas. 
Nevertheless, this mode demonstrates a high quality factor, leading to a high transmittance.

The situation changes entirely for incident waves with moderate and large powers [see Figs.~\ref{fig:fig3}(b) and (c), respectively]. 
In these cases, the resonant localized defect mode Eq.~(\ref{eq:dmode}) is progressively disintegrated due to the fact that the associated line width increases strongly. 
The latter is now dictated by the nonlinear losses $\Im\,\Omega(\Ind)$ occurring at the defect resonator, rather than the radiative losses associated with the edges of the CRMW. 
As a result, the $Q$ factor of the resonant mode is deteriorated leading to an underdamped-to-overdamped transition. 
The critical incident intensity for which this transition occurs can be estimated by imposing a critical coupling criterion:
\begin{eqnarray}\label{eq:critcoupling}
\Gamma_\textrm{rad} &\sim& \Gamma_\textrm{bulk};\\
\Gamma_\textrm{rad} &\sim& |\psi_{n_L}|^2+ |\psi_{n_R}|^2= \frac{2}{\xi} e^{-\frac{|(N-1)/2|}{\xi}},\\
\Gamma_\textrm{bulk} &\sim& \Im(\nu_{n_D})|\psi_{n_D}|^2=\Im(\nu_{n_D})/\xi\,,
\end{eqnarray}
where $n_L = 1$ and $n_R = 17$. 
Both $\Gamma_\textrm{rad}$ and $\Gamma_\textrm{bulk}$ are evaluated using first-order perturbation theory. 
We stress that the destruction of the localized resonant-defect mode also signifies the mitigation of any electrical breakdown effects for the diode. 

In Figs.~\ref{fig:fig3}(d)-(f), we show the scattering resonant profiles (solid lines) for the same incident powers as for the upper panel, still assuming $\gamma=0$ but with $z_0 \ne z_1$ taking the experimentally measured values. 
Thus $\Omega(\Ind) = z_0-z_1\ne 0$ for low incident powers, but is still very small. 
Nevertheless, this affects the staggered form of the resonant field profile [see Fig.~\ref{fig:fig3}(d)] and partially distorts the resonant-defect mode. 
As the incident power increases, the resonant mode initially remains intact [see Fig.~\ref{fig:fig3}(e)]. 
However, for even higher incident powers, the defect mode quickly deteriorates and does not show any remnants of localization around the defect resonator. 
Rather, it decays exponentially from the incident edge of the CRMW [see Fig.~\ref{fig:fig3}(f)]. 
The same behavior is observed in the presence of bulk losses (dashed lines) from the resonators ($\gamma \ne 0$). 
The latter describes nicely the experimental profiles (dotted-dashed lines) shown in the same figure. 
The transformation of the defect resonant mode Eq.~(\ref{eq:dmode}) to an exponentially decaying field signifies the onset of a suppressed transmittance -- the latter 
being roughly proportional to $\left|\frac {\psi_N}{\psi_1}\right|^2\sim \exp(-N/\xi)$ with a characteristic decay length $\xi$. 
Indeed, we show below that the underdamping-to-overdamping transition is the dominant mechanism responsible for the suppression of the transmittance of the resonant defect mode. 
An expected, nonlinear resonance detuning at high powers is certainly helpful but not necessary for the suppression of the transmittance, since the defect mode is already destroyed long before the detuning becomes relevant for the current analysis \cite{Note_arXjeo20,arXjeo20}.

\begin{figure}
	\centering
	\includegraphics[width=\columnwidth]{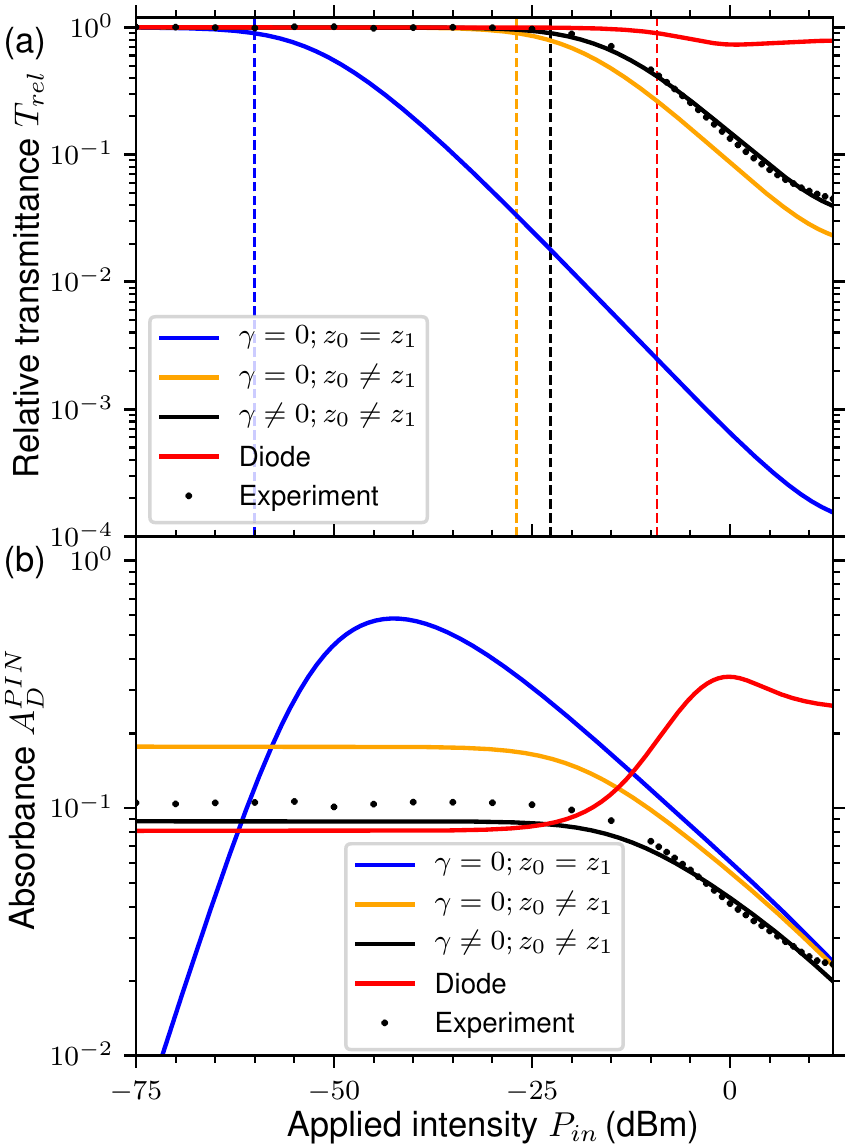}
	\caption{\label{fig:fig4}
		(a) transmittance and (b) absorbance of the photonic RP versus the intensity $P_{in}$ of the incident electromagnetic signal.
		The solid lines correspond to theoretical results, while the black dots indicate our measurements from the setup of Fig.~\ref{fig:fig1}(b).
		Additionally the transmittance for the single diode is shown (red line).
	}
\end{figure}

To quantify this expectation for the performance of our RP, we present in Fig.~\ref{fig:fig4}(a) the relative transmittance $T_{\rm rel}=T / T_{\rm low}$ ($T_{\rm low}$ is the transmission at small incident powers) versus the incident power. 
For each incident power, we trace the associated resonance frequency and extract the corresponding transmittance $T$ 
-- thus making sure that we evaluate its maximum (peak) value all the time \cite{Note_arXjeo20}.
Three different cases are considered:
(a) zero bulk losses from the resonators $\gamma=0$ and zero linear losses of the diode $z_0=z_1$ (ideal design; see blue line);
(b) zero bulk losses $\gamma=0$ but nonzero linear losses due to the presence of the diode $z_0\ne z_1$ (orange line); and
(c) nonzero bulk losses $\gamma\ne 0$ and nonzero linear losses due to the presence of the diode $z_0\ne z_1$ (corresponding to our experimental setup; see black line). 
In the same figure, we present the measurements of our RP (black dots). 
For high incident powers, the experimental values of $T_{\rm rel}$ indicate a saturation, trend which is a consequence of the rather shallow transmission floor $T$ within the band gap of the SSH structure arising mainly from the width of the resonances at the two band edges.
This limitation can be further improved by increasing the number of dimers on each side of the topological defect or by increasing the quality factor of the individual resonators.
Finally, for comparison, we additionally report the transmittance for the single diode (red line). 

In Fig.~\ref{fig:fig4}, we also indicate by vertical dashed lines the LT of each of these cases, which is identified as the value of the incident power for which $T_{\rm rel}\approx 90\%$.
In fact, the LT can be approximated using the critical-coupling criterion Eq.~(\ref{eq:critcoupling}).
In the ideal scenario, the Ohmic losses are dominated by the nonlinear loss at the defect site $\Im(\Omega)$ and thus $\Gamma_{\textrm{bulk}} \sim \Im(\Omega(\Ind))/\xi$.
Using Eq.~(\ref{eq:dmode}), we can express $\Ind$ in terms of $I_{n_L}$, which results in an equation to solve for $I_{n_L} \sim \textrm{LT}$:
\begin{equation}\label{eq:genlt}
2\chi \sim \Im(\Omega(\chi^{-1}I_{n_L}))\,,
\end{equation}
where $\chi = e^{-\frac{|(N-1)/2|}{\xi}}$.
Eq.~(\ref{eq:genlt}) holds for a general form of nonlinearity.
For illustration, let us consider the limit of weak incident power.
Expanding the nonlinearity to first order, $\Im(\Omega(\Ind)) = (z_0 - z_1) + \Im(z_1 \alpha) \Ind$.
The resulting LT is given by
\begin{equation}\label{eq:lt}
I_{n_L}
\sim \frac{2 \chi^2 - (z_0 - z_1) \chi}{\Im(z_1 \alpha)}
\end{equation}
Note that $\chi$ describes the growth of the scattering field intensity near the defect site.
With a larger system size, a higher field intensity peak -- and hence lower LT -- can be achieved.
In agreement with our expectations, the LT decreases exponentially with an increase in $N$ [see Eq.~(\ref{eq:lt})].
Using the backward-transfer-matrix approach, we verify that the LT decreases substantially with a growing system size in the ideal scenario. 

The excellent agreement between our theoretical results (black line) and the measurements is a direct confirmation of the accuracy of our nonlinear coupled-mode modeling scheme and verifies the proposed RP mechanism as described.
It is obvious that our RP has a dramatic limiting effect, providing a 3-orders-of-magnitude suppression of the transmittance.
As opposed to the stand-alone diode, we see that when the diode is incorporated in the CRMW array, the LT is reduced by approximately 15\,dBm, while the theoretical upper bound (corresponding to an "ideal" design--see above) allows for a LT that is approximately 50\,dBm smaller than that of the single diode.

Let us finally discuss the power of the proposed RP in terms of self-protection against high incident power.
We have already established that the RP prevents electrical breakdown of the diode component due to the destruction of the localized defect resonant mode at high incident powers.
However, there is another threat, associated with excessively absorbed energy of the incident radiation by the \emph{sensitive} diode.
This amount of energy turns to heat and will eventually lead to a destruction of the diode.
Using Eqs.~\eqref{eq:Mn}, we can show that the absorbance at resonator $n_D$ is expressed in terms of the field intensity at the defect resonator as follows:
\begin{eqnarray}\label{eq:eq6}
A_D &=&A_D^{res}+A_D^{PIN};\\
A_D^{res}&=&2\gamma\frac{|\psi_{n_D}/r_0|^2}{\nu_g},\\
A_D^{PIN}&=&2Im(\Omega(\Ind))\frac{|\psi_{n_D}/r_0|^2}{\nu_g} \,,
\end{eqnarray}
where $\nu_g=\partial\omega(k)/\partial k$ is the group velocity at the leads.
In Eq.~\eqref{eq:eq6}, the term $A_D^{res}$ is associated with the bulk losses of the ceramic resonator, while $A_D^{PIN}$ describes the energy absorbed by the PIN diode and is the main topic of interest for us.
From Eq.~\eqref{eq:eq6}, we calculate $A_D^{PIN}$ for the three scenarios mentioned above.
The results are presented in Fig.~\ref{fig:fig4}(b) together with the measurements of the photonic RP (black dots).
A nice agreement is again observed between our measurements and the corresponding theoretical curve (black line).
In the case of high incident powers, all scenarios associated with the photonic RP demonstrate a more than one-order-of-magnitude reduced absorption with respect to the stand-alone diode (red line).
In fact, in the ideal setup, we find that the absorbance for high powers could be further suppressed by increasing the system size $N$, which implies an increase in the damage threshold.
We conclude, therefore, that the coupling of a standard diode with a CRMW array can mitigate its destruction by overheating from high-power incident radiation.
In this limit, the proposed RP reflects almost all of the energy, leading to a large dynamic range. 

Although at this paper, we investigate the limiting properties of the SSH photonic structure of Fig.~\ref{fig:fig1} against CW signals, we expect a similar high-quality limiting performance for pulse signals as well. 
In this latter case, one needs to make sure that the carrier frequency of the pulse is located inside the band gap and has a bandwidth that confines the signal within the gap. 
A potential concern might be associated with the recovery time of the nonlinear element (diode) to respond to a fast pulse. 
Fast diodes with quick recovery time of the order of tens of nanoseconds are, however, available \cite{RP1,RP2} and can be used instead of the SMS\,7630-079LF Schottky Diode that we incorporate in the current design. 
The operational power of the limiter can be further configured by increasing or decreasing the number of dimers of the SSH circuit (see previous discussion), by adjusting the distance between the defect resonator and the diode or the diameter of the short circuiting [see inset in fig.~\ref{fig:fig1}(c)] or simply by using diodes with other specifications. 
These latter adjustments will calibrate the critical power values for which the nonlinear losses (due to the diode) are activated, allowing us a further control of the LT (see 
discussion above) of the proposed photonic circuit.

\section{CONCLUSIONS}
Utilizing the framework of topological photonics, we devise and implement a RP consisting of a bipartite CRMW array coupled inductively at a defect site with a PIN diode with lossy nonlinearity.
For low incident powers, the symmetry-protected defect resonance mode provides robust high transmittance, which reduces appreciably when the power of the incident signal increases beyond a critical value.
This LT is reduced considerably with respect to the corresponding LT of the single PIN diode, as used in conventional designs.
At the same time, the damage threshold of the photonic RP is increased via the reduction in absorbance by at least one order of magnitude as compared to a typical RP associated with the stand-alone PIN diode.
The defining parameters of the CMRW setup, such as the couplings to the leads or dc voltages applied to the diode, can in principle be controlled externally, e.g., electronically, giving rise to an adaptable design with a controllable LT.
The proposed scheme is scalable, can be printed in CMOS, and can be used as a protection element for a variety of rf transceiver setups.

More generally, this work provides inspiration to find applications of other modern concepts from photonic scattering and transport theory. 
For instance, coherent perfect absorbers \cite{cho10,pic19} could prove useful for the dumper channel, and signals could also be directed utilizing unidirectional invisibility \cite{lin11}, direction-independent wave propagation \cite{jin18}, and helicity-dependent transport \cite{zha19}. 
Furthermore, all of these effects could be further enhanced by considering their nonlinear extensions.

\section*{ACKNOWLEDGMENTS}
T.K. and D. H. J. acknowledge partial support from the Office of Naval Research (ONR) N00014-16-1-2803 and from the Defense Advanced Research Project Agency (DARPA) Nascent Light-Matter Interactions (NLM) program via Grant No.~HR00111820042.


\begin{thebibliography}{37}
\bibitem [{Not()}]{Note_arXjeo20}%
	\BibitemOpen
	\href@noop {} {}\bibinfo {note} {The shape and behavior of a nonlinear
		topologically protected defect mode supported in an SSH system is
		investigated in detail \cite{arXjeo20}}\BibitemShut {NoStop}%
\end{thebibliography}
\end{document}